# Comparing dynamic correlation lengths from an approximation to the four-point dynamic susceptibility and from the picosecond vibrational dynamics.

D. Fragiadakis, R. Casalini, and C.M. Roland
Naval Research Lab, Chemistry Division, Code 6120, Washington, DC 20375-5342

*(August 31, 2011)*

**ABSTRACT.** Recently a new approach to the determination of dynamic correlation lengths, $\xi$, for supercooled liquids, based on the properties of the slow (picosecond) vibrational dynamics, was carried out [L. Hong, V.N. Novikov, and A.P. Sokolov, Phys. Rev. E **83**, 061508 (2011)]. Although these vibrational measurements are typically conducted well below the glass transition temperature, $T_g$, the assumption is that the structure of the liquid is frozen at $T_g$, so that the $\xi$ characterize dynamic heterogeneity in the supercooled liquid state. We compare $\xi$ from this method to values calculated using an approximation to the four-point dynamic susceptibility. For 26 different materials we find good correlation between the two measures; moreover, the pressure dependences are consistent within the large experimental error. However, $\xi$ from Boson peak measurements above $T_g$ have a different, and unrealistic, temperature dependence.

______________________________________________

A universal feature of liquids approaching their glass transition is an increasing length scale for the correlations between molecular rearrangements. This property, and the associated spatial heterogeneity of the motions, are inherent to the many-body dynamics and must be accounted for by theories of the glass transition [1]. This requires reliable data quantifying the dynamic correlation length, $\xi$. An immediate difficulty is unambiguously defining the distance over which motions are correlated, given the fractal nature of the heterogeneous dynamics. Various approaches have been applied to the determination of $\xi$ [2,3]. The maximum value of the dynamic susceptibility, $\chi_4$, yields the number of dynamically correlated molecules, $N_c$ [4]; however, since $\chi_4$ involves both spatial and temporal correlations, it cannot be obtained from linear relaxation measurements without assumptions. Berthier et al. [5,6] proposed an approximation in terms of the temperature derivative of the experimentally accessible, two-point dynamic correlation function, $\chi_T(t)$

$$\chi_T(t) = \frac{\partial C(t)}{\partial T} \leq T^{-1}\left(\frac{\Delta c_P}{k}\chi_4(t)\right)^{1/2} \tag{1}$$

where $\Delta c_P$ is the isobaric heat capacity change at $T_g$, and $k$ is the Boltzmann constant. The lower bound on the number of dynamically correlated molecules is

$$N_c \geq \frac{k}{\Delta c_P}T^2[\max \chi_T(t)]^2 \tag{2}$$

$\max \chi_T(t)$ representing the maximum value of $\chi_T(t)$ for any state point. If the dielectric loss has the Kohlrausch form with a stretch exponent, $\beta$, that varies only weakly with temperature (which is true around $T_g$), eq. (2) can be written as [7]

$$N_c \geq \frac{k}{\Delta c_P}\left(\frac{\beta}{e}\right)^2\left(\frac{d\ln\tau_\alpha}{d\ln T}\right)^2 \tag{3}$$

where e is Euler's number.

Confidence in the validity of $\chi_T$ as an approximation to $\chi_4$ draws from several observations:

(i) The number of particles involved in dynamic heterogeneities near $T_g$ has a reasonable magnitude (*ca.* 100) [6,7,8,9,10].

(ii) The increase in $N_c$ is strongest at temperatures approaching $T_g$, in accord with the divergence of the relaxation time near $T_g$.

(iii) Although the $\chi_T$ approximation becomes less accurate at temperatures above $T_g$, due to an increasing contribution from (neglected) density fluctuations, $N_c$ from $\chi_T$ decreases to negligible values at sufficiently high temperature, consistent with molecular motions becoming non-cooperative [8].

(iv) Molecular dynamics simulations (mds) comparing the approximate $N_c$ from $\chi_T$ and exact values from $\chi_4$ agree over the supercooled regime [9,11]. And close to the glass transition, experimental $N_c$ from $\chi_T$ are in agreement with the correlation length determined by multi-dimensional NMR [5].

An alternative means to quantify $\xi$ recently developed by Hong et al. [12,13,14] is based on the fact that dynamic heterogeneities are manifested in the fast (picosecond) dynamics [15,16]. Dynamic heterogeneity in the glassy state has been observed by mds: mobile particles, defined as those having large vibrational amplitudes, have a propensity to be surrounded by mobile particles, and less mobile particles tend to have a disproportionate number of immobile neighbors [17]. Although these clusters are more compact than those observed above $T_g$ [17], the cessation of relaxation below the glass transition freezes in structural inhomogeneities, resulting in an excess density of states and responsible for the Boson peak in the inelastic scattering of amorphous solids. Spatial fluctuations of the local elastic constants induce corresponding fluctuations in the vibrational amplitudes (mean squared displacements), with the length scale of the fluctuations, $\xi_{BP}$, inversely proportional to the frequency of the Boson peak, $v_B$ [18]

$$\xi_{BP} = Sc/v_B \qquad (4)$$

where $c$ is the transverse sound speed and the constant $S$ is about equal to unity. The implication is that structural heterogeneities intrinsic to the glassy state can be identified with the dynamic heterogeneities detected in the supercooled liquid by relaxation measurements [12,13,14].

Experimental confirmation that $\xi_{BP}$ is indeed a measure of the correlation length relevant to the dynamics of supercooled liquids has thus far been limited to comparisons involving four materials (glycerol, sorbitol, OTP, and polyvinylacetate) [13] that had previously been studied by multi-dimensional NMR. The NMR method enables an exact determination of $\chi_4$, but only in the vicinity of $T_g$ [19]. In this report we compare $\xi_{BP}$ to dynamic correlation lengths determined from $\chi_T(t)$ for twenty-six materials, including inorganic glass-formers and molecular and polymeric organic liquids. To do this we convert the number of correlating molecules calculated from eqs (3) to a length scale using [10,20]

$$\xi_\chi = \left(N_c V_m\right)^{1/3} \qquad (5)$$

where $V_m$ is the molecular volume. Eq (5) ignores any string-like character [4] of the cooperative volume and provides, for a given $N_c$, an upper bound for the correlation length.

In Figure 1 we plot the $\xi_{BP}$ reported in refs. [12,13] versus $\xi_\chi$, using for the latter literature results [21] for the quantities in eq (3) and the equations of state used to calculate $V_m$. All the correlation lengths in Fig. 1 correspond to $T_g$: $\xi_\chi$ was determined from quantities measured at $T_g$, and although the Boson peak was measured for the glass, the structure is assumed to have become fixed at the glass transition, so that the $\xi_{BP}$ correspond to the values at $T_g$. The two measures of the dynamic correlation length are seen to be in good agreement, Pearson's correlation coefficient equal to 0.84. Even though the $N_c$ from eq (3) is a lower bound, the $\xi_\chi$ in Fig. 1 are typically about 40% larger than the $\xi_{BP}$; however, this is in part an artifact of using eq

(5) to convert the number of correlating molecules to a length scale. Assuming a spherical volume [22], $\xi_\chi$ and $\xi_{BP}$ would be more nearly equal in magnitude.

Another comparison of the two measures of correlation lengths is their change for a given material with temperature and pressure. Hong et al. [12] reported $\xi_{BP}$ versus pressure for seven compounds; these results, normalized by the ambient pressure values, are reproduced in Figure 2 for the three materials (OTP, polyisoprene, and polymethylphenylsiloxane) for which $\xi_\chi$ could be calculated. For pressures up to 1.4 GPa, the changes in $\xi_{BP}$ do not exceed 30%; however, as noted by Hong et al. [12], the data are contradictory concerning whether the correlation length increases or decreases with pressure.

We showed previously that when isochronal superpositioning [1] is valid, the change in $N_c$ with pressure is given by [8]

$$N_c \sim \left(1 + \gamma \alpha_P T\right)^2 / \Delta c_p \qquad (6)$$

in which $\alpha_P$ is the isobaric thermal expansion coefficient and $\gamma$ the scaling exponent in the equation relating the structural relaxation time, $\tau_\alpha$, to temperature and density, $\rho$ [1,23]

$$\tau_\alpha = f(T\rho^{-\gamma}) \qquad (7)$$

$\gamma$ is a material constant, and since the change with pressure of the heat capacity difference at $T_g$ between the liquid and glassy states is negligible [24,25], the variation of $N_c(T_g)$ in eq. (6) with pressure is governed by changes in the quantity $(1+ \gamma \alpha_P T_g)$. At ambient pressure the product $\alpha_P T_g$ is a near universal constant (the Boyer-Spencer rule [26]), but this quantity decreases for a given material with pressure. This change can be assessed from the pressure-dependence of the isobaric fragility [8]

$$m_p = m_V \left(1 + \gamma \alpha_P T_g\right) \qquad (8)$$

where $m_V$ is the (pressure-independent [27]) isochoric fragility. (Note from eqs (6) and (8) that the dynamic correlation length cannot correlate with the fragility.) Experimentally for van der Waals liquids, $m_p$ has been found to decrease about 10% over the range from ambient to 0.5 GPa [27], implying about the same variation in $\xi_{BP}$ over the larger pressure range in Fig. 2. This variation is within the scatter for the data, although the correlation length for normal liquids is unambiguously a decreasing function of pressure, notwithstanding the behavior of $\xi_{BP}$ for OTP.

Also plotted in Fig. 2 are $\xi_\chi$ calculated using eqs (5) and (6) for these three materials, at pressures for which the required data were available; again we normalize by the ambient pressure value for each. The magnitudes of the changes with pressure are similar to those for $\xi_{BP}$, but $\xi_\chi$ always has the expected inverse dependence on $P$. One significant difference between the determinations of $\xi_{BP}$ and $\xi_\chi$ is that while data for the latter were obtained directly at $T_g(P)$, for $\xi_{BP}$ the Boson peak was actually measured at 140K, following application of pressure at room temperature and subsequent cooling to induce the transition to the glassy state. This method relies on the assumptions that the structure of the glass is fixed, and there is no significant effect of temperature *per se*; therefore, the dynamic correlation length measured in the glassy state represents the value expected for the supercooled liquid at $T_g$.

In Figure 3 the two correlation lengths for OTP and glycerol at ambient pressure are plotted as a function of temperature normalized by $T_g$. $\xi_\chi$ decreases with increasing temperature, with the steepest change being around the glass transition (and steeper for OTP than for glycerol, the former the more fragile liquid [28,29]). In contrast, $\xi_{BP}$ shows negligible change with temperature, behavior that cannot be reconciled with our understanding of the glass transition. Indeed, the fact that the dynamic correlation length depends primarily on $\tau_\alpha$ [1,8] means that $\xi$ must change markedly in the vicinity of $T_g$. Note that whereas the $\xi_{BP}$ in Figs. 1 and 2 were measured below $T_g$, thus corresponding to values at the glass transition, the results in Fig. 3 were based on measurements of the liquids; that is, at $T > T_g$. It is unclear whether the structure is actually liquid-like, since the time scale (ps) of the vibrational measurements is shorter than $\tau_\alpha$. The dependence of $\xi$ on $\tau_\alpha$ also is at odds with the suggested correlation of the former with the activation volume ($\Delta V^{\#} = RT\partial \ln(\tau)/\partial P|_T$) [12]. As pointed out by Hong et al. [12], one possibility is that the prefactor $S$ in eq. (4) is temperature dependent.

In summary, correlation lengths calculated as the ratio of the transverse sound speed and the Boson peak frequency, both measured in the glassy state, correspond well to values of the dynamic correlation length at $T_g$ determined using an approximation to the four-point dynamic susceptibility. However, $\xi_{BP}$ measured for the liquid above $T_g$ have an unrealistic *T*-dependence, suggesting a disconnect between low frequency structural relaxation and THz vibrational measurements on the liquid state.

We thank Vladimir Novikov for kindly providing data from ref. [12]. DF acknowledges a National Research Council post-doctoral fellowship. This work was supported by the Office of Naval Research.

# FIGURE CAPTIONS

**Figure 1.** Dynamic correlation length defined as the ratio of the transverse sound speed to the Boson peak frequency measured at 140K versus the same quantity at $T_g$ calculated from the approximation to the 4-point dynamic susceptibility. The fitted line has a slope = 0.68, with a Pearson correlation coefficient as indicated.

**Figure 2.** Dynamic correlation length calculated from eq. (3) (open symbols) and eq.(4) (filled symbols) and normalized by the value at ambient pressure, as a function of pressure for *o*-terphenyl (circles), 1,4-polyisoprene (squares), and polymethyphenylsiloxane (diamonds). The $\xi_\chi$ data are at the pressure-dependent $T_g$, while for $\xi_{BP}$ the samples were pressurized at room temperature and then quenched to 140K, which is below the glass transition temperature in each case.

**Figure 3.** Dynamic correlation length normalized by the value at $T_g$ calculated using eq. (3) (open symbols) and eq.(4) (filled symbols) plotted versus temperature over $T_g$, for glycerol (circles) and *o*-terphenyl (squares).

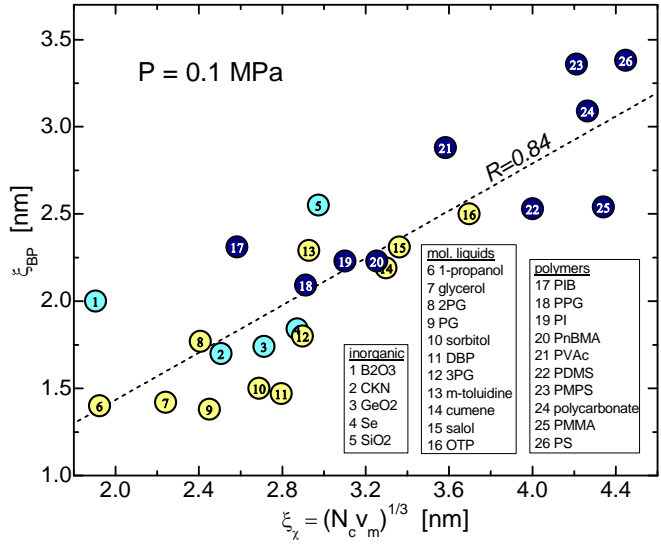

Figure 1

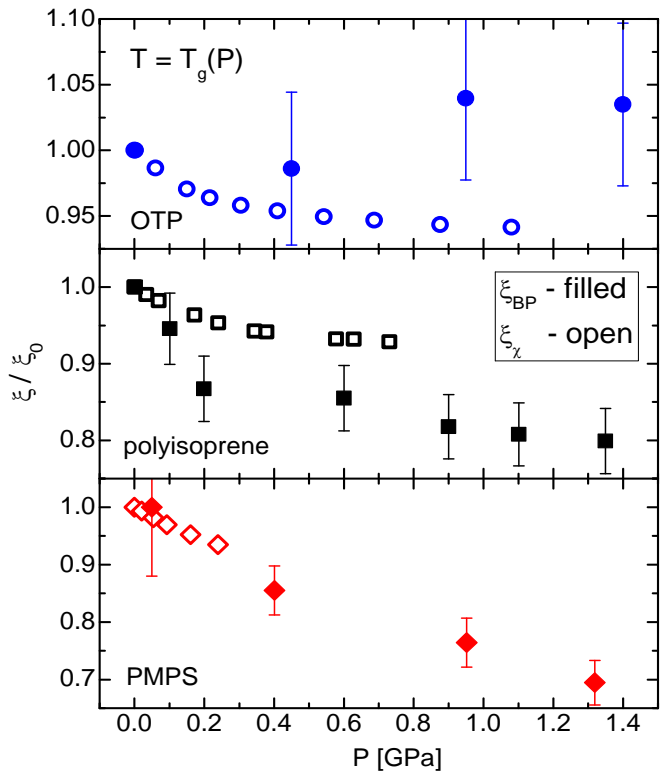

Figure 2

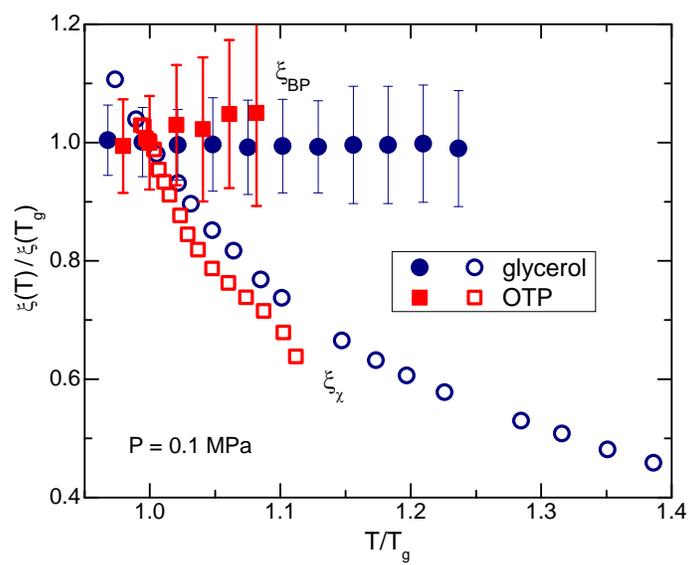

Figure 3